\documentclass[prd,aps,twocolumn,a4paper,floatfix]{revtex4}

\usepackage{graphicx,psfrag}
\usepackage{mathrsfs}
\usepackage{amsmath,amsfonts,amssymb} 
\usepackage{times}

\def\p{\partial}

\usepackage{color}
\usepackage{float}
\definecolor{cyan}{rgb}{0,0.9,0.9}
\definecolor{orange}{rgb}{0.9,0.5,0}
\definecolor{magenta}{rgb}{1,0,1}
\definecolor{purple}{rgb}{0.8,0.4,0.8}

\begin{document}

\title{%
  The trumpet solution from spherical gravitational collapse with
  puncture gauges}

\author{Marcus \surname{Thierfelder}$^1$}
\affiliation{$^1$Theoretical Physics Institute, University of 
  Jena, 07743 Jena, Germany}
\author{Sebastiano \surname{Bernuzzi}$^1$}
\affiliation{$^1$Theoretical Physics Institute, University of 
  Jena, 07743 Jena, Germany}
\author{David \surname{Hilditch}$^1$} 
\affiliation{$^1$Theoretical Physics Institute, University of 
  Jena, 07743 Jena, Germany}
\author{Bernd \surname{Br\"ugmann}$^1$}
\affiliation{$^1$Theoretical Physics Institute, University of 
  Jena, 07743 Jena, Germany}
\author{Luciano \surname{Rezzolla}$^2$}
\affiliation{$^2$Max-Planck-Institut f\"ur Gravitationsphysik, 
  Albert-Einstein-Institut, Golm, Germany}

\date{\today}

\begin{abstract}
  We investigate the stationary end-state obtained by evolving a
  collapsing spherical star with the gauges routinely adopted to study
  puncture black holes. We compare the end-state of the collapse with
  the trumpet solution found in the evolution of a single wormhole
  slice and show that the two solutions closely agree. We demonstrate
  that the agreement is caused by the use of the Gamma-driver shift
  condition, which allows the matter to fall inwards into a region of
  spacetime that is not resolved by the numerical grid, and which
  simultaneously finds the stationary coordinates of the trumpet outside
  the matter.
\end{abstract}


\maketitle


\section{Introduction}
\label{sec:Introduction}

The possibility of obtaining stable evolutions of relativistic compact
objects within the framework of numerical relativity relies crucially
on the choice of suitable coordinate gauge conditions. In
particular, the well-known \emph{puncture} approach relies on the
gauge conditions to deal with explicit coordinate singularities
present in the computational domain. These gauge conditions are built
upon the ``1+log'' lapse condition~\cite{Bona:1994dr} and the
``Gamma-driver'' shift condition~\cite{Alcubierre:2002kk}, and
allow the coordinate singularity (puncture) to be advected across the
grid by the shift~\cite{Baker:2005vv, Campanelli:2005dd,Hannam:2006vv,vanMeter:2006vi,Pollney:2007ss}. Hereafter we
will refer to these gauge conditions as the \textit{puncture
  gauges.}
  
It has been shown~\cite{Hannam:2006vv,Hannam:2006xw,Brown:2007tb,Garfinkle:2007yt,Hannam:2008sg,Dennison:2010wd} that, for a single 
puncture without linear momentum or spin, the wormhole topology
of the puncture initial data~\cite{Brandt:1997tf} ceases to be
resolved by the mesh in the code. Instead, the mesh approaches an
asymptotically cylindrical stationary, solution, \textit{i.e.,} the
\textit{trumpet} solution. The puncture gauges can thus be viewed
as a sort of ``natural excision'' which squeezes the singularity into
an unresolved region of the computational
domain. It is nowadays commonly
used in binary black hole simulations by many groups
following~\cite{Baker:2005vv,Campanelli:2005dd}.

The puncture gauges have furthermore been tested in a number of codes
solving also the equations of relativistic
hydrodynamics~\cite{Duez:2005sf,Baiotti:2006wm,Shibata:2006bs,Loffler:2006nu,Faber:2007dv,Baiotti:2008ra}. It
has been found 
effective in handling the black hole which forms in the collapse of a
neutron star~\cite{Baiotti:2006wm,Baiotti:2007np,Montero:2008yx}, and in
binary systems in general (see
e.g.~\cite{Shibata:2006bs,Loffler:2006nu,Etienne:2007jg,Baiotti:2008ra,Giacomazzo:2009mp}
and references 
therein). In~\cite{Baiotti:2006wm}, in particular, it was first shown
with three-dimensional (3D) simulations that when using the
puncture gauges, no special treatment beyond standard artificial
dissipation for the metric variables is necessary to follow stably the
black-hole formation and evolution.

In this paper we show that the numerical evolution, with the puncture
gauge, of a collapsing spherical star approaches the trumpet solution
at late times. This result is not trivial for a number of
reasons. First, because in the collapsing spacetime, and in
particular in the portion filled by matter, there is no time-like
Killing vector and it is not clear how a stationary end-state can be
found.  Second, at the continuum level, the two spacetimes are clearly
different since, for example, slices in the two foliations have a 
different topology. Even restricting our attention to the vacuum 
region of the collapsing spacetime and assuming a Killing slice 
compatible with the ``1+log'' lapse condition, it is not obvious that 
the resulting slice will be exactly that reported 
in~\cite{Hannam:2006vv}. Finally, the collapsing matter inside 
the apparent horizon is observed to disappear but it is not clear
whether this reflects a physical or a numerical behavior.

We therefore seek answers to the following questions: (i)~How
well do the two endstates agree?  (ii)~Where does the matter go
after the apparent horizon is formed?  (iii)~How do the gauges
find a stationary slicing of a spacetime without an explicit time-like
Killing vector? In order to address these questions we evolve the
spacetime of a star collapsing to a black hole using a ``1+log''
slicing condition and two variants of the Gamma-driver 
shift condition. The different results are then compared {\it
  quantitatively} with the trumpet solution, and spacetime diagrams
are constructed to follow the motion of the matter. Overall we find
that the two spacetimes tend to a common stationary solution at late
times and that this agreement is primarily caused by the
shift condition, which pushes grid-points away from regions of high
curvature, preventing the matter from being resolved on the
numerical mesh.

The plan of the paper is as follows. In Sect.~\ref{sec:Method} we
briefly review our numerical methods, presenting the numerical
results and their interpretation in
Sect.~\ref{sec:Numerical_Results}. Our conclusions are summarized in
Sect.~\ref{sec:Conclusion}, while we dedicate the
Appendix~\ref{app:SpaceTimeDiag} to a description of our method for
constructing spacetime diagrams for spherical spacetimes.

\section{Numerical Setup}
\label{sec:Method}

We perform numerical simulations both in explicit spherical symmetry
by means of the one-dimensional (1D) code described
in~\cite{Bernuzzi:2009ex} and in 3D by means of {\sc
  BAMmatter}~\cite{Thierfelder-prep}.  The latter is developed
extending the hydrodynamics solver of the spherical code into the {\sc
  BAM} code~\cite{Brugmann:2008zz}.  More specifically, we solve the
full set of Einstein equations in the $3+1$ formalism coupled to
general relativistic hydrodynamics (GRHD). In our
spherically-symmetric simulations we adopt both the BSSNOK and
Z4c formulations of the Einstein field equations
(see~\cite{Bernuzzi:2009ex} for details), while in 3D we use only
the BSSNOK formulation.

As mentioned above, our gauges are built upon the ``1+log''
condition~\cite{Bona:1994dr} for the lapse $\alpha$ and the Gamma-driver
condition for the shift $\beta^i$~\cite{Alcubierre:2002kk} written in
the form
\begin{align}
\label{eq:lapse_1log}
\p_t\alpha   &= \beta^i\alpha_{,i}-\alpha^2\mu_L\hat{K},\\
\label{eq:beta_Gdriv}
\p_t\beta^i  &= \mu_S\tilde{\Gamma}^i-\eta
\beta^i+\beta^j\beta^i{}_{,j}\,,
\end{align}
where we always choose $\mu_L=2/\alpha$, with either 
$\mu_S=1$ or $\mu_S=\alpha^2$ and $\eta=2/M$. The Gamma-driver 
condition~\eqref{eq:beta_Gdriv} may be obtained as the first integral 
of what is implemented in most numerical-relativity codes. A 
comparison of the behavior of the condition~\eqref{eq:beta_Gdriv} 
with the standard form in the evolution of puncture black holes
is presented in~\cite{vanMeter:2006vi}.

For the evolution of the matter we adopt the Valencia
flux-conservative formulation of the equations of
  GRHD~\cite{Banyuls:1997,Font:2001ew,Baiotti:2004wn} for a perfect
  fluid. A property of flux-conservative formulations is that they
preserve (exactly in the continuum) certain integral quantities
such as the total rest-mass of the fluid, $M_0$, or its momentum
and energy density. The equations of state (EoS) employed to
describe the fluid are the ideal-gas or the polytropic
EoS~\cite{Font:2001ew}: no significant differences were found when
using one or the other.

In both codes the numerical evolution in time
relies on the method-of-lines with Runge-Kutta integrators and
finite difference approximations. The GRHD equations are solved by
means of a high-resolution shock-capturing (HRSC) scheme based on the Local
Lax-Friedrich (LLF) central scheme~\cite{Kurganov:2000} combined with 
Convex-Essentially-Non-Oscillatory (CENO)~\cite{Liu:1998} 
reconstruction~\cite{Bernuzzi:2009ex,Thierfelder-prep}.

In analogy with~\cite{Baiotti:2006wm} and in contrast
to~\cite{Bernuzzi:2009ex}, we do not find it necessary to excise the
hydrodynamical variables in the collapse evolutions. Artificial
Kreiss-Oliger dissipation is added in the evolution of the metric
fields following the standard procedure~\cite{Brugmann:2008zz}.
Instead, we find it important to set the GRHD eigenvalues to zero if
unphysical values are computed 
(``eigenvalue excision'').  Note that this is compatible with the use
of the LLF central scheme, which requires only an estimate of the
local speed. Unphysical values can be produced in a neighborhood of
the center of the collapse, due to numerical errors.  In 3D
simulations we also found it important to set a ceiling on the Lorentz
factor $W$ in order to prevent the code from crashing
after the formation of the apparent horizon. In particular $W$ is 
set to the ceiling value $W_{\rm ceil}=10^{10}$ when (and only when) the 
velocity becomes larger than the speed of light.
Simulations employing the excision of the hydrodynamical quantities
were also performed. In this case, the matter variables are set to the
atmosphere value in a small region well inside the apparent
horizon. As expected, no differences were found. In the following
we focus on simulations without excision.

The initial stellar model is an unstable spherical configuration widely
used in the literature,
\textit{e.g.,}~\cite{Font:2001ew}. Adopting units in which
$c=G=M_{\odot}=1$, the configuration chosen has central rest-mass
density $\rho_c=7.993\times10^{-3}$, gravitational (ADM) mass $M=1.448$
($M_0=1.535$) and circumferential radius $R=5.838$ (isotropic coordinate
radius $r_R=4.268$). The collapse is triggered imposing a negative
velocity perturbation which is larger than the truncation
error~\cite{Font:2001ew}. For the vacuum simulations we use standard
puncture data~\cite{Brandt:1997tf} in isotropic coordinates and with
an ADM mass which is the same as that of the star.

The simulations in 1D were performed on a grid with uniform spacing
with resolutions $\Delta_{\rm r} = 0.02, 0.01, 0.005$. The
simulations in 3D, on the other hand, were performed imposing an octant
symmetry on a cell-centered Cartesian grid with 8 fixed mesh refinement
levels, with the resolutions of finest level given by $\Delta_{\rm
  xyz}= 0.05 , 0.03125 , 0.025$, and where the resolution doubles
from one level to the next.

\section{Results}
\label{sec:Numerical_Results}

\begin{figure}[t]
  \begin{center}    
\includegraphics[width=0.45\textwidth]{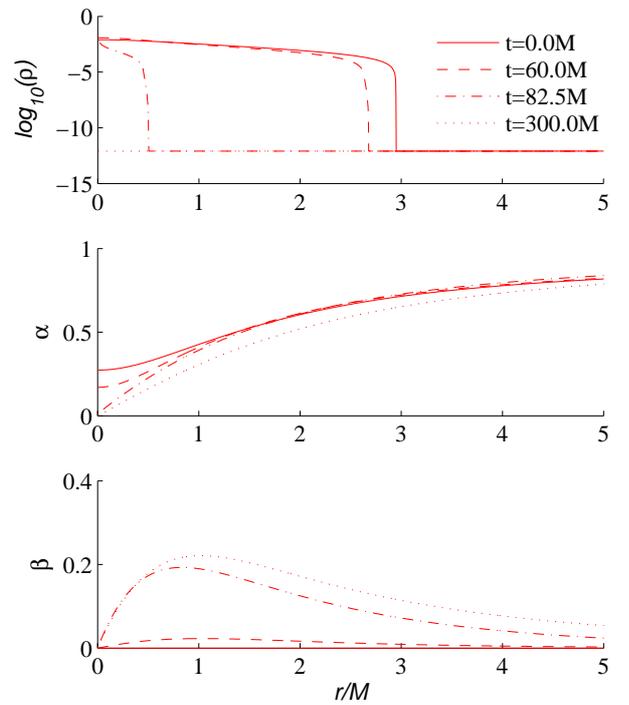} 
    \caption{\label{fig:dynamics} Radial profiles of the rest-mass
      density $\rho$ (top panel), of the lapse function $\alpha$
      (middle panel), and of the shift $\beta\equiv \beta^r$ (bottom
      panel) at some representative times of the evolution. Note that
      at $t\sim50\,M$ an apparent horizon is first found and that by
      $t=300\,M$ the amount of matter on the final time slice is
      essentially that in the atmosphere.}
  \end{center}
\end{figure}

We next describe the results of the different numerical
simulations. Before that, however, it is useful to recall some of
the basic properties of the evolutions that from the puncture
spacetime lead to the trumpet solution or that from a stellar
spacetime lead to a black-hole solution.

\paragraph*{Puncture spacetime.} The evolution of the puncture
spacetime is described in Ref.~\cite{Hannam:2006vv,Hannam:2008sg}. The
initial Schwarzschild slice in isotropic coordinate evolves, driven by
the puncture gauges, to the stationary trumpet solution.  Approximate
stationarity is achieved within an evolution time of about $60M$,
depending on the details of the gauge choice. Asymptotically, there is
a coordinate singularity at the puncture point at $r=0$, which
corresponds to a sphere with Schwarzschild radius $R_0\approx1.3M$.
The trumpet slice extends from $R_0$ to spatial infinity, $i^0$, while
the wormhole slice of the initial data reaches from the outer $i^0$ to
an inner spatial infinity. 
Conceptually it is important to distinguish between the analytical
solution and the numerical solution on a discrete grid, and also
between the wormhole slice of the analytical evolution and the trumpet
slice which is only reached asymptotically.  The evolution starts with
a wormhole slice which, due to continuity of the analytical solution,
remains a wormhole slice. It asymptotes to a trumpet slice in the
sense that the region inside $R_0$ has a coordinate size that tends
towards zero. Numerically, this inner region is effectively excised
once its coordinate size drops below the grid spacing. Assuming that
there is no grid point at the puncture itself, after a short evolution
time the numerical grid only has grid points outside $R_0$, that is
only the trumpet part of the initial wormhole is represented.
Anticipating the discussion of matter, the key difference is that the
initial data with matter lives neither on a wormhole nor a trumpet
slice, since the matter ``covers'' the inner region of the slice.

\paragraph*{Collapsing spacetime.} The dynamics of a collapsing 
unstable star in general relativity has been discussed in great detail
in a number of papers and we refer the interested reader to
Refs.~\cite{Font:2001ew,Baiotti:2004wn,Baiotti:2006wm} for some of
the most recent work. For the test-case considered here ($\mu_S=1$), the
introduction of the perturbation is sufficient to drive the star over
the stability threshold and induce its gravitational collapse. As a
result, matter essentially freely falls towards the center, leading to
an exponential growth of the rest-mass density and a related rapid variation of the metric
functions. This is shown in the top panel of Fig.~\ref{fig:dynamics},
which exhibits the radial profile of the rest-mass density at some
representative times. The ``singularity-avoiding'' properties of the
``1+log'' slicing condition drive the lapse function to very small
values near the origin. This is shown in the middle panel of
Fig.~\ref{fig:dynamics}, while the bottom one refers to the evolution
of the shift. Note that as the matter rushes towards the center, the
shift still succeeds in arresting the motion of the radial coordinates
outside the matter distribution (ignoring the atmosphere), thereby
preventing slice stretching (in the sense of stretching of the spatial
coordinates) as for the black hole evolutions.
At about $t\sim50\, M$, an apparent horizon forms indicating
unambiguously the presence of a black hole. This time represents the
time when a first comparison between the two spacetimes can in
principle be made. We also note that because the coordinate radius of
the apparent horizon is initially $r_{\text AH}\lesssim 2\,M$, part of
the matter is outside of it, but it is then rapidly accreted.

As highlighted in~\cite{Baiotti:2006wm}, the gauge conditions in
Eqs.~\eqref{eq:lapse_1log}--\eqref{eq:beta_Gdriv} allow us to follow
the subsequent evolution without having to excise either the
hydrodynamical or the gravitational field variable. The conservation
of the rest-mass is very good up to the horizon formation: $\Delta
M_0(t)/M_0(0)\leq 0.05\%$. After the formation of the apparent
horizon, the matter inside is observed to disappear from the numerical
grid, so that by $t= 300\,M$ the amount of matter on the slice is that
in the atmosphere, $M_0(300) \sim 10^{-6}\,M_0(0)$. As we will discuss 
in Sect.~\ref{sec:matterinside}, a suitable change in the value of 
$\mu_S$ for the shift condition can avoid this behavior.

As a side remark we note that in our 1D simulations we observe a
better behavior of the Z4c formulation of the Einstein equation
with respect to the BSSNOK one in terms of constraint violation and 
long-term stability. In particular, at $t=2500\,M$, the L2
norm of the Hamiltonian constraint is about $4$ orders of magnitude
lower with Z4c. Furthermore, at $t=2500\,M$, the irreducible mass of
the final black hole in the Z4c evolutions is within the numerical
error of the ADM mass of the initial data. In contrast, the
BSSNOK simulations display a significant deviation in the irreducible
mass of the black hole after~$t=300\,M$. In view of this, in what
follows the 1D results we will present refer exclusively to those
obtained for the Z4c formulation.

\begin{figure}[t]
  \begin{center}    
\includegraphics[width=0.45\textwidth]{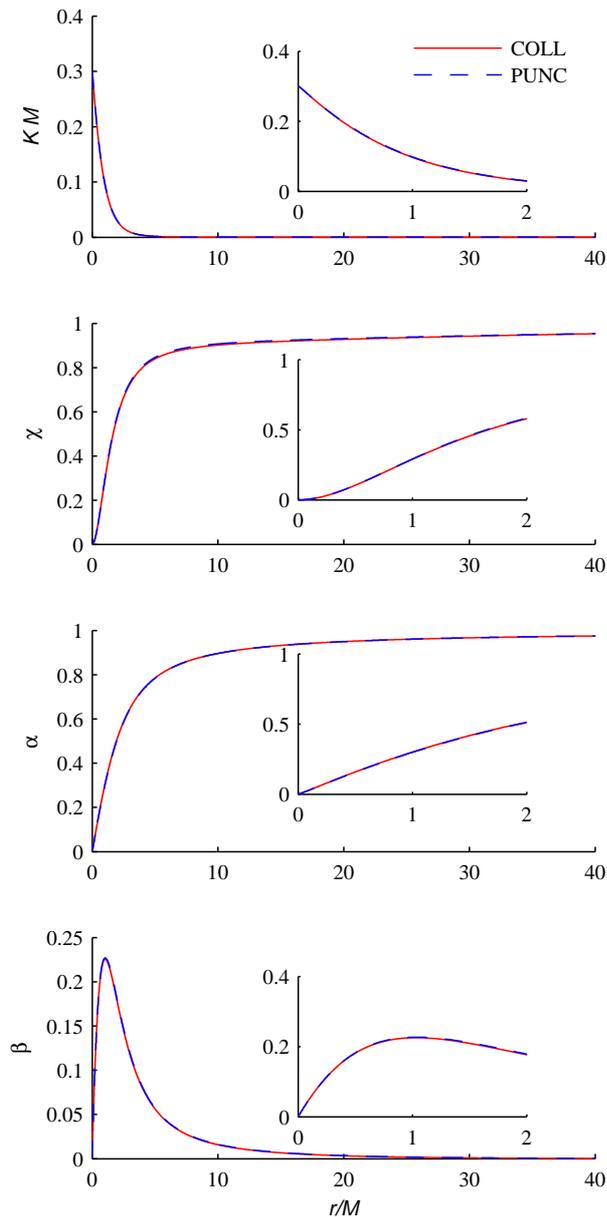} 
    \caption{\label{fig:fields1d_at_t100} Radial profiles of
        representative field variables at time $t=300\, M$ from the
      1D simulations of the puncture and the collapsing star. From top
      to bottom: trace of extrinsic curvature ($K$), conformal factor
      ($\chi$), lapse ($\alpha$) and shift ($\beta\equiv\beta^r$).}
  \end{center}
\end{figure}

\begin{figure}[t]
  \begin{center}
\includegraphics[width=0.45\textwidth]{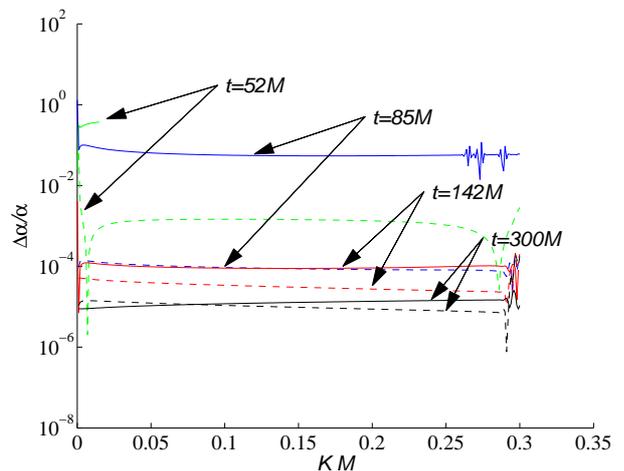}
    \caption{\label{fig:alpha_K_1d} Relative difference $\Delta
      \alpha/\alpha \equiv \alpha_{\rm punc}/\alpha_{\rm coll}-1$,
      between the lapse of the puncture evolution, $\alpha_{\rm
        punc}$, and that of the collapsing star, $\alpha_{\rm coll}$,
      as a function of the trace of the extrinsic curvature, $K$, at
      different coordinate times. For a fair comparison, data at the
      same $K$ are found by interpolation. The dashed lines refer to
      the relative differences between the analytic result and
      the puncture evolution. By construction, for early times this
      difference is much smaller than the difference between puncture
      and matter simulations.
      }
  \end{center}
\end{figure}

\subsection{Agreement of the end-states}

In this Section we investigate the asymptotic slice reached at
the end of the evolution by the collapsing star and compare it
with the corresponding trumpet one. This is shown in
Fig.~\ref{fig:fields1d_at_t100}, which reports several metric
fields at time $t=300\, M$. This time is well after the apparent
horizon is first found (\textit{i.e.,}~$t_{\textrm{AH}}\sim 50\,M$),
and represents in both cases a time when the solution has become
essentially stationary. From top to bottom the different panels
refer respectively to: the trace of extrinsic curvature ($K$), the
conformal factor ($\chi$), the lapse ($\alpha$) and the shift
($\beta^r$). It is quite apparent that at the selected time the two
spacetimes are extremely similar and it is difficult to distinguish
the solutions two by a visual inspection.

To obtain a more quantitative estimate of the differences we have
computed the behavior of the fields in the collapsing spacetime near
the origin and obtained the following fitting functions for the
stationary solution for $r \ll 1$:
\begin{eqnarray}
\label{eq:K0}
K\,M &\sim&  0.30 - 0.37 \, \left(\frac{r}{M}\right)\, ,\\
\label{eq:chi0}
\chi &\sim& 1.22 \, \left(\frac{r}{M}\right)^{2.0 } \, ,\\
\label{eq:alpha0}
\alpha &\sim& 0.54 \, \left(\frac{r}{M}\right)^{1.09} \, .
\end{eqnarray}
The fits for \eqref{eq:chi0} and \eqref{eq:alpha0} contain the
exponent of $r$ as a fitting parameter, but not for \eqref{eq:K0}.
The result agrees well with the corresponding expressions
in~\cite{Brugmann:2009gc}. In particular, the non-integer exponent for
the lapse in \eqref{eq:alpha0} is very close to the analytic
expression for the trumpet solution in isotropic coordinates, which
has exponent $1.091$ \cite{Brugmann:2009gc}. This is interesting since
the numerical coordinates are not isotropic, although one could argue
that close to the puncture the isotropy of the initial data is
maintained during the evolution.

This result is confirmed when considered also in a
coordinate-independent manner. Following the
prescription suggested in~\cite{Hannam:2006vv}, we analyze
the dependence of the lapse versus the extrinsic curvature
and report in Fig.~\ref{fig:alpha_K_1d} the differences at
different times. More specifically, we show with solid lines the
relative difference $\Delta \alpha/\alpha \equiv \alpha_{\rm
  punc}/\alpha_{\rm coll}-1$, between the lapse of the puncture
evolution, $\alpha_{\rm punc}$, and that of the collapsing star,
$\alpha_{\rm coll}$, either when the apparent horizon has just
formed ($t\sim 50\,M$) or when the solutions have reached a
stationary stage ($t\sim 300\,M$).

It is clear that the relative difference decreases in time and, at
time $t=300\, M$, it is below~$0.1\%$.  By performing convergence
tests we have also determined that the numerical errors are at the
same level as the~$0.1\%$ disagreement. Also reported in
Fig.~\ref{fig:alpha_K_1d} with dashed lines is the relative
difference between the puncture data and the analytic solution for
the trumpet solution~\cite{Brugmann:2009gc,Hannam:2008sg}
\begin{eqnarray}
\label{eq:trump}
K &=& \beta \alpha'(R) / 2\\
&=& \frac{\sqrt{{2}/{R(\alpha)}+\alpha ^2-1} \left(4 R(\alpha)
  \alpha ^2-4 R(\alpha)+6\right)}{2 R(\alpha) \left(R(\alpha) \alpha
  ^2-2 R(\alpha) \alpha -R(\alpha)+2\right)} \nonumber \ ,
\end{eqnarray}
where $R$ is the Schwarzschild radius, $\alpha'$ is the derivative 
 of $\alpha$ with respect to $R$ and we set $M=1$.  As expected,
the relative difference is in this case much smaller initially,
but it becomes comparable with the one computed for the collapsing
spacetime at later times. 

\begin{figure}[t]
  \begin{center}
\includegraphics[width=0.45\textwidth]{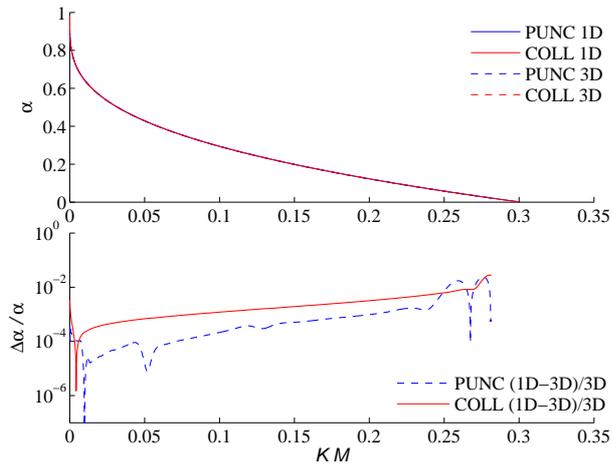}
    \caption{\label{fig:alpha_K_3d1d} Slicing from 1D and 3D
      simulations. Top panel: the lapse ($\alpha$) is plotted versus the
      trace of the extrinsic curvature $K$ at time $t=300\, M$ for
      the collapse in 1D (red solid line) and the puncture
      evolution (blue solid line) spacetimes, and the collapse
      in 3D (red dashed line) and collapse (blue dashed line)
      spacetimes. Bottom panel: relative differences. Note that 1D
      results, computed on a finer grid, are interpolated for the
      comparison.}
  \end{center}
\end{figure}

It is possible to model the behavior of the data from the collapsing
spacetime near the origin and at $t=300\, M$ as
\begin{equation}
\label{eq:Kalpha0}
K\,M \sim 0.30 - 0.92 \, \alpha\ .
\end{equation}

Similarly, a Taylor expansion of Eq.~\eqref{eq:trump} around
$\alpha=0$ (\textit{i.e.,} for the values of the lapse near the
puncture) can be performed by using the implicit function
$R(\alpha)$ in~\cite{Brugmann:2009gc}, yielding
\begin{eqnarray}
K(\alpha) &=& 0.300937 -0.930916 \alpha + O\left(\alpha^2\right) \ ,
\end{eqnarray}
which closely agrees with Eq.~\eqref{eq:Kalpha0}.

Finally, we note that we did not find significant differences
between the 1D and the 3D results for the collapsing
spacetime. This is summarized in Fig.~\ref{fig:alpha_K_3d1d},
where the top panel shows $\alpha$ versus $K$ for the 1D and the
3D data at time $t=300\,M$: they are visually
indistinguishable. Similarly, the bottom panel shows that the relative
differences between the 1D and the 3D data are generically 
below $2\%$.

\subsection{Where does the matter go once inside the horizon?}
\label{sec:matterinside}

\begin{figure}[t]
  \begin{center}    
\includegraphics[width=0.45\textwidth]{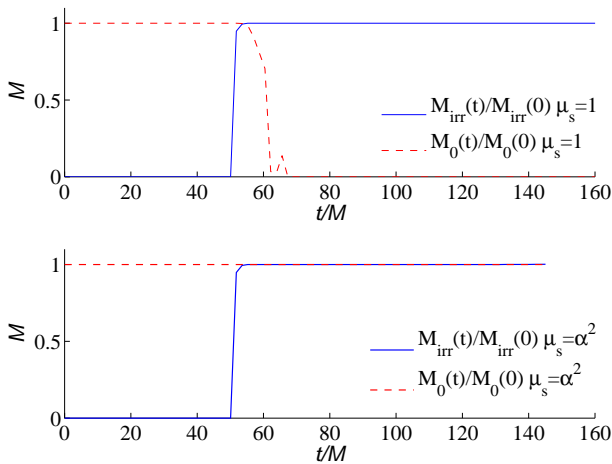}\\
    \caption{\label{fig:masses} Evolution of total rest-mass,
      $M_0(t)$, (red dashed lines) and of the irreducible mass, $M_{\rm
        irr}(t)$, (blue solid lines) normalized respectively to
      the initial value and to the ADM mass. The top panel refers to a
      simulation with gauge speed $\mu_s=1$, and the bottom panel to
      evolutions with gauge speed $\mu_s=\alpha^2$. }
  \end{center}
\end{figure}

As discussed above, the numerical evidence obtained when using the
Gamma-driver shift condition with $\mu_S=1$ is that the matter
inside the apparent horizon is progressively ``dissipated'' (see
also~\cite{Baiotti:2006wm} where this was first discussed).
This behavior is clearly displayed in the upper panel of
Fig.~\ref{fig:masses}, in which the total rest mass normalized to the
initial value is plotted
in time (dashed red line). Note that at about $10M$ after the
formation of the apparent horizon the 
 normalized rest mass drops smoothly to zero except for a small
bump near $t=65M$ generated by numerical errors inside the horizon.
Also shown is the irreducible mass of the black hole normalized to the
ADM mass (solid blue line),
which is obviously zero before the apparent horizon is found. This
behavior may appear puzzling since it is geometrically clear that
the matter cannot leave the foliation (see \textit{e.g.,~}Fig.~6.11
of~\cite{Wald:1984rg}). In addition, since we are using HRSC
schemes, our numerical methods should be sufficiently robust, for
reasonable spatial resolutions, to handle extremely large gradients
as the matter piles up in the collapse. Indeed, we have
experimented with various reconstruction algorithms and resolutions
and found that both affect slightly the rate and the initial time of
the disappearance of the fluid. More specifically, more
dissipative schemes result in an earlier disappearance of the
matter, while higher resolutions can delay it. However, neither
improvement prevents the disappearance of the matter from the grid.

In order to establish whether this behavior is instead due to an
excessive stretching of the spatial coordinates generated by the Gamma-driver
shift condition, we have performed the same simulations using either
a shift gauge speed $\mu_S=\alpha^2$, or simply setting
$\beta^i=0$. We recall that this is possible since the ``1+log''
lapse condition is a pure slicing condition, so that the foliation
is unaffected by a change of radial gauge (see for example
~\cite{Garfinkle:2007yt,Alcubierre:2002iq}).
  
\begin{figure}[t]
  \begin{center}    
\includegraphics[width=0.38\textwidth]{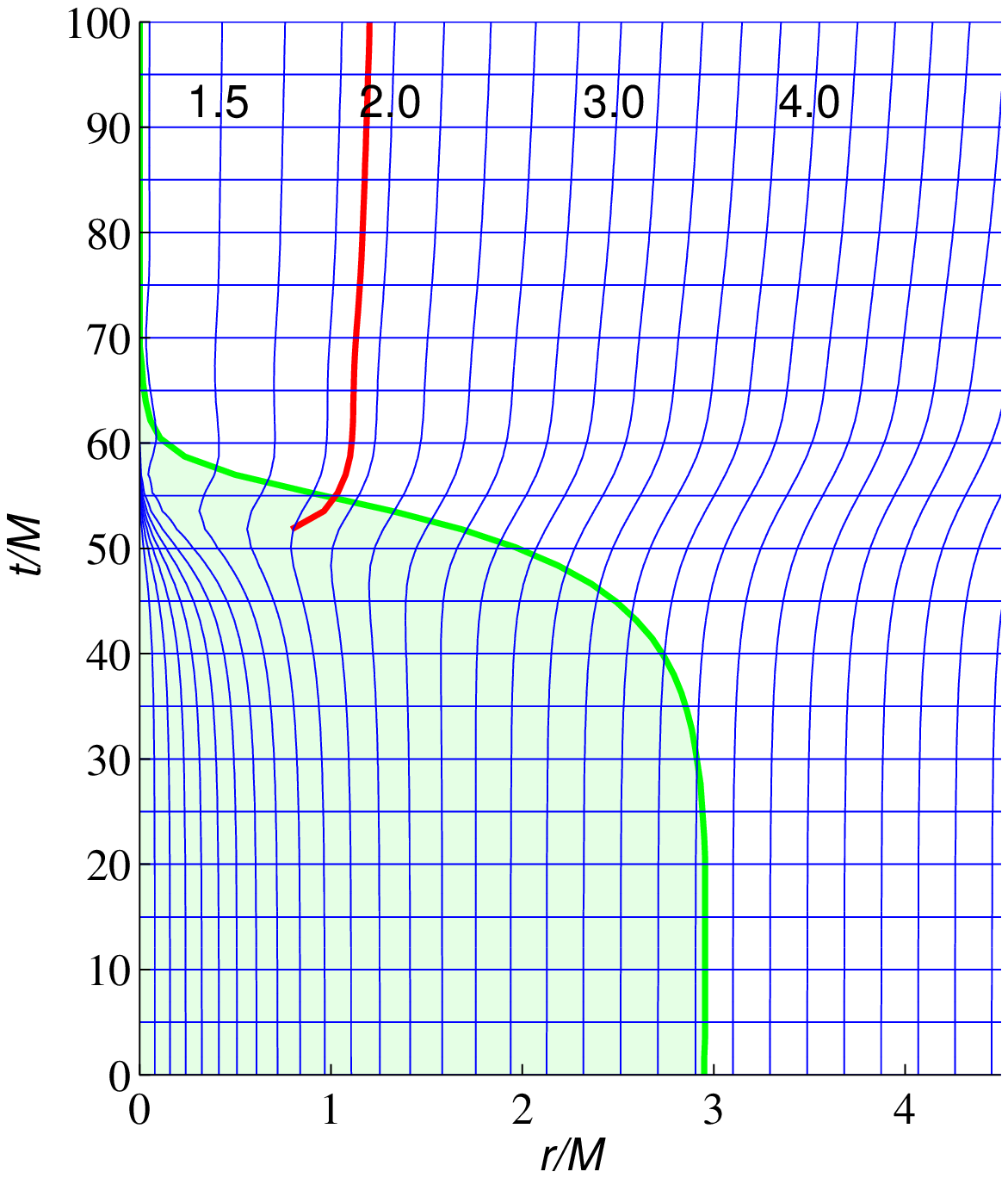}
\includegraphics[width=0.38\textwidth]{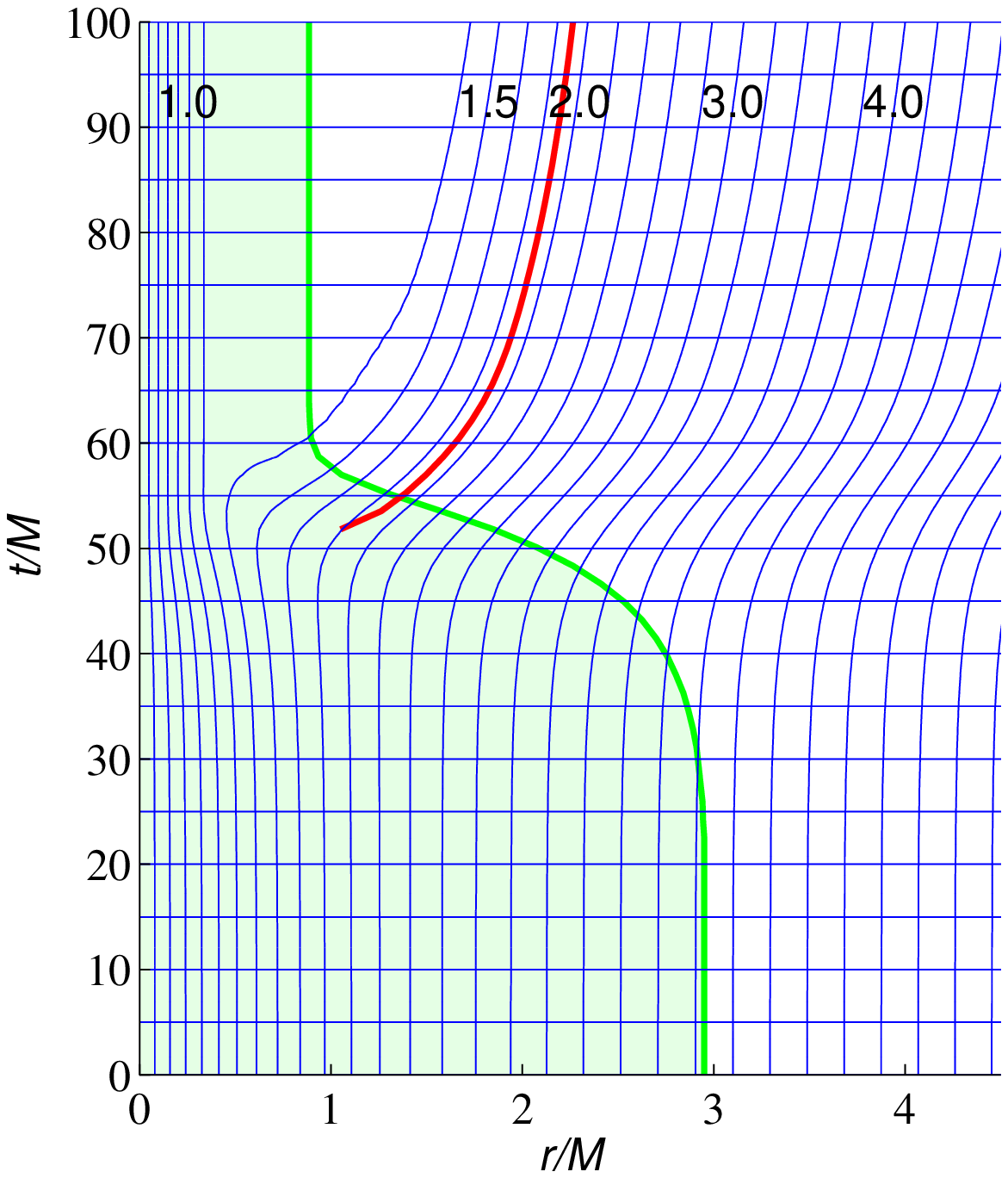}
     \caption{\label{fig:spctmdiag_rt}
       Spacetime diagram of the
       collapsing star. Data are from 1D simulations with gauge speed
       $\mu_S=1$ (top panel) and $\mu_S=\alpha^2$ (bottom panel). 
       The horizontal blue lines are lines of constant coordinate time.
       The thick red line denotes the apparent horizon. The
       vertical blue lines are lines of constant Schwarzschild 
       radius $R$ which values are on top of the lines. The shaded green
       area bounded by the thick green lines shows the region of the 
       matter.}
  \end{center}
\end{figure}

The results of these tests are shown in the bottom panel of Fig.~\ref{fig:masses},
where we report with the evolution of the total rest-mass (dashed red
lines) and of the normalized irreducible mass (solid blue lines) when
$\mu_S=\alpha^2$. Clearly, while the behavior of the irreducible mass is independent of the choice for
$\mu_S$, that of the rest-mass is not. When using the $\mu_S=\alpha^2$
radial gauge, in fact, the matter remains on the numerical grid, so
that the rest-mass is conserved well beyond the formation of the
apparent horizon.  The cause of this difference must therefore to be
attributed to the large stretching of the spatial coordinates with
$\mu_S=1$.  Specifically, we identify two effects.  Firstly matter 
falls inside the innermost grid point as the inner boundary is 
effectively an outflow boundary for the matter; $M_0$ starts to drop 
to the atmosphere value approximately $10\,M$ before the stellar surface
passes through the innermost gridpoint. At this time, the areal radius
of the innermost gridpoint grows rapidly to approximately $1.9\,M$
(see also Fig.~\ref{fig:spctmdiag_rt} later).
Secondly the effective resolution in areal radius near the time of the
stellar surface passing through the innermost gridpoint is
approximately an order of magnitude lower than in the initial
data. The stretching is thus so large that the matter ``percolates''
through the grid as the numerical methods are not able to reproduce
its steep gradients. As a result, the spacetime ``empties'' itself and
this explains the very good match with the trumpet solution reported
in Fig.~\ref{fig:alpha_K_1d}. Conversely, when $\mu_S=\alpha^2$, the
radial gauge does not distort the grid significantly, allowing for an
excellent conservation of the rest-mass on the grid. Note that the
latter is not the result of the rather high spatial resolution, but it
is simply the result of the coordinate time ``freezing'' induced by
the collapsed lapse function. Of course, because the spacetime is not
able to remove its matter content, the match with the trumpet solution
is in this case much worse and $\Delta \alpha/\alpha \sim 10^{-1}$ for
$K\,M\gtrsim 0.21$.

Not surprisingly, much of what we discussed so far for $\mu_S=\alpha^2$
applies also when considering $\beta^i=0$. However, the same
dynamics that leads a $\beta^i=0$ shift condition to fail in a
curved spacetime, is responsible for the late-time failure of the
simulations having $\mu_S=\alpha^2$ as gauge speed. As a result, a
Gamma-driver shift condition with $\mu_S=1$ appears to be the
most robust choice for all those situations in which a compact fluid
object may collapse to a black hole.

\begin{figure}[t]
  \begin{center}    
\includegraphics[width=0.48\textwidth]{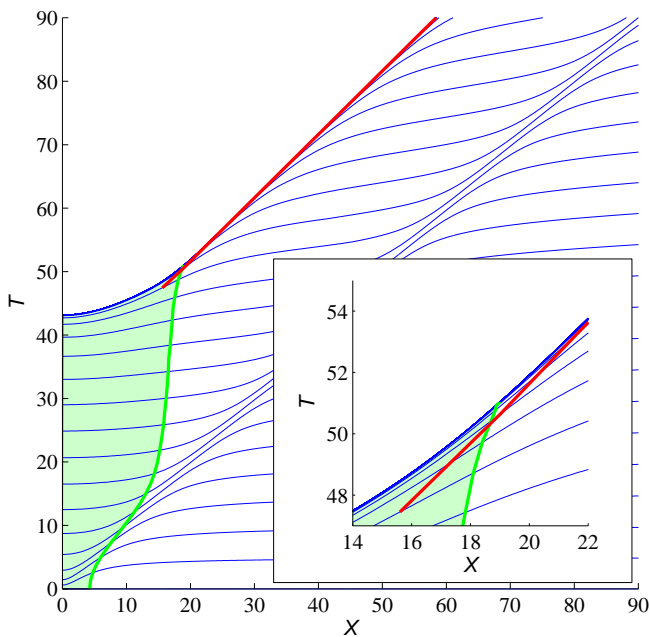}
     \caption{\label{fig:spctmdiag}
        Causal spacetime diagram of the collapsing star. 
        Data are from 1D simulations with gauge speed
        $\mu_S=\alpha^2$. The coordinate axes are constructed so that
        the light speeds have magnitude one.
        The horizontal blue lines are lines of constant coordinate time.
        The thick red line denotes the apparent horizon.
        The shaded green area bounded by the thick green lines shows
        the region of the matter.
}
  \end{center}
\end{figure}

Finally we show in Figs.~\ref{fig:spctmdiag_rt} and
\ref{fig:spctmdiag} the spacetime diagrams as built from the numerical 
data. The axes in Fig.~\ref{fig:spctmdiag_rt} refer to the
time and spatial coordinates used respectively in the simulations 
with ~$\mu_S=1$ (upper panel) and
~$\mu_S=\alpha^2$ (bottom panel). The shaded green area corresponds to
the region of the spacetime covered by matter, vertical lines are line
of constant Schwarzschild radius while horizontal lines are lines of
constant coordinate time. The thick red line emerging after
$t\sim50\,M$ is the apparent horizon. Fig.~\ref{fig:spctmdiag_rt} can be compared with
other numerically-generated diagrams of collapsing
spacetimes obtained either in other gauges~\cite{ST85b}, or with
excision techniques~\cite{Baiotti:2004wn}, or for
puncture evolutions of single black holes~\cite{Hannam:2008sg}. The figure 
clearly shows how the matter is ``squeezed''
from the numerical grid when $\mu_S=1$, while it remains on the grid when
$\mu_S=\alpha^2$. In both cases an apparent horizon is found. The
comparison of the lines of constant Schwarzschild radius in the two
panels highlights the stretching of the spatial coordinate
discussed before.

Figure~\ref{fig:spctmdiag} displays the causal spacetime diagram of
the collapse. Similarly to Fig.~\ref{fig:spctmdiag_rt} the thick red
line denotes the apparent horizon, the thick green line the surface of the
collapsing star and the shaded area shows the region filled by  
the matter. The coordinate $(X,T)$ are constructed so that the light
speeds have magnitude one, with light cones opening at $45$ degree in the
diagram. The method employed for the construction is discussed
in Appendix~\ref{app:SpaceTimeDiag}. Because easier to 
use, the data employed to construct the diagram refers to a 1D 
simulation with~$\mu_S=\alpha^2$, but in principle we expect that a very similar 
diagram would be produced with~$\mu_S=1$. 
At early times the causal spacetime diagram clearly shows
the extremely high speed of the stellar surface and the causal separation
between the points initially belonging to the 
exterior and those that are causally connected to the collapse 
(see the gauge-wave propagating at the speed of light).
At late times it is evident that the collapse of the lapse prevents the
slices from evolving forward into  
the singularity; furthermore the inset demonstrates that there is always
some numerical data inside the horizon.

\section{Conclusion}
\label{sec:Conclusion}


By comparing the numerical evolution of a single puncture with
that of a collapsing star when using the same puncture gauges,
we have shown numerically that the two spacetimes tend to the same
trumpet solution at late times, a possibility already conjectured
in~\cite{Baiotti:2006wm}. Let us explicitly address the 
questions raised in the introduction: (i) In the domain covered by 
the numerical coordinates, at late times, the spacetimes agree 
within the precision of our calculations. (ii) The apparently bizarre 
agreement, which is not possible at the continuum level, is caused 
primarily by the Gamma-driver shift condition, which stretches the 
numerical spatial grid. The matter is rapidly forced inside the 
innermost grid point, preventing it from being resolved and 
effectively removing it from the spatial slice. Thus in the domain 
covered by the numerical coordinates the spacetimes agree, solving 
the apparent contradiction. (iii) Since the matter is lost inside 
the innermost gridpoint, at late times, on the numerical grid the 
evolution is simply that of the vacuum Schwarzschild solution.

We also note that if the matter in the slice is sufficiently
compact and in the vacuum region and one has a Killing slicing
which is compatible with the ``1+log'' gauge, then in the vacuum
region the slice must agree with the original ``1+log''
trumpet~\cite{O'Murchada:private}. Our analysis also demonstrates
that, whilst the puncture gauges with $\mu_S=1$ results in robust
numerical evolutions in a collapse scenario, by construction 
the resulting coordinates are not appropriate for a detailed study 
of the dynamics of the matter near the singularity.

\appendix

\section{The construction of spacetime diagrams}
\label{app:SpaceTimeDiag}

In this appendix we describe our approach to the construction of
spacetime diagrams in spherical symmetry. The approach is based on
that of ~\cite{Ortiz:2010}. The most general spherical line element
can be written as
\begin{align}
\textrm{d}s^2=-\alpha^2\textrm{d}t^2
+\gamma_{rr}(\textrm{d}r+\beta^r\textrm{d}t)^2
+\gamma_{\theta\theta}\textrm{d}\Omega^2,
\end{align}
where $\textrm{d}\Omega^2=\textrm{d}\theta^2+\sin^2\theta \,\textrm{d}\phi^2$ 
is the standard metric on the two-sphere. The incoming and 
outgoing radial light speeds are given by 
\begin{align}
c_{\pm} = -\beta^r\pm\frac{\alpha}{\sqrt{\gamma_{rr}}}.
\end{align}
Ingoing and outgoing null coordinates $(u,v)$ satisfy the 
equations of motion
\begin{align}
\p_tu = - c_{+}\p_ru, &\qquad \p_tv = - c_{-}\p_rv,
\end{align}
in coordinates $(t,r,\theta,\phi)$. Assuming that near the outer
boundary $r_*$ space is almost flat, we arrive at the boundary 
condition
\begin{align}
v(t,r_*)&= 2t-u.
\end{align}
We introduce the scalar fields
\begin{align}
X=\frac{1}{2}(v-u),&\qquad
T=\frac{1}{2}(u+v).
\end{align}
They satisfy the equations of motion
\begin{align}
\p_tX=\beta^r\p_rX+\frac{\alpha}{\sqrt{\gamma_{rr}}}\p_rT,&\qquad \!\!\!\!
\p_tT=\beta^r\p_rT+\frac{\alpha}{\sqrt{\gamma_{rr}}}\p_rX.
\end{align}
and are naturally adjusted to the causal structure of the 
spacetime. In terms of $(T,X,\theta,\phi)$ 
the spherical line-element becomes
\begin{align}
\textrm{d}s^2= \psi^4(-\textrm{d}T^2+\textrm{d}X^2)+
\gamma_{\theta\theta}\textrm{d}\Omega^2,
\end{align}
where $\psi^4$ is a conformal factor determined by 
\begin{align}
\psi^4&= \frac{\gamma_{rr}}{(\p_rX)^2-(\p_rT)^2}.
\end{align}
To construct spacetime diagrams we simply evolve the fields
$(u,v)$. As initial data we choose $u=-r$ and $v=r$. There are
potential problems associated with the breakdown of the coordinate
chart, but we will not concern ourselves with those issues here.  From
our computations we observed however that the method requires further
development (\textit{e.g.,~}a treatment of the fields at the horizon)
to construct spacetime diagrams for every numerically generated
spacetime (cf.\ Sec.~III~B of~\cite{Hannam:2008sg}). In particular we
found the construction of causal spacetime diagrams with $\mu_S=1$ 
troublesome, perhaps because more sophisticated boundary conditions 
are required.

\bigskip
\begin{acknowledgments}

We thank Niall O'Murchadha and Luca Baiotti for discussions.  This
work was supported in part by DFG grant SFB/Transregio~7
``Gravitational Wave Astronomy'' and by ``CompStar'', a Research
Networking Programme of the ESF.

\end{acknowledgments}



\end{document}